# Phase stability and structural temperature dependence in sodium niobate: A high resolution powder neutron diffraction study

S. K. Mishra[1], R. Mittal[1], V. Yu. Pomjakushin[2] and S.L. Chaplot[1]

[1]Solid State Physics Division, Bhabha Atomic Research Centre, Mumbai 400085, India.
[2]Laboratory for Neutron Scattering, ETH Zurich and Paul Scherrer Institut, CH-5232 Villigen PSI, Switzerland.

## Abstract

We report investigation of structural phase transitions in technologically important material sodium niobate as a function of temperature on heating over 300-1075 K. Our high resolution powder neutron diffraction data show variety of structural phase transitions ranging from non-polar antiferrodistortive to ferroelectric and antiferroelectric in nature. Discontinuous jump in lattice parameters is found only at 633 K that indicates that the transition of orthorhombic antiferroelectric P (space group *Pbcm*) to R (space group *Pbnm*) phase is first order in nature, while other successive phase transitions are of second order. New superlattice reflections appear at 680 K (R phase) and 770 K (S phase) that could be indexed using an intermediate long-period modulated orthorhombic structure whose lattice parameter along <001> direction is 3 and 6 times that of the $CaTiO_3$-like *Pbnm* structure respectively. The correlation of superlattice reflections with the phonon instability is discussed. The critical exponent (β) for the second order tetragonal to cubic phase transition at 950 K, corresponds to a value $β≈^1/_3$, as obtained from the temperature variation of order parameters (tilt angle and intensity of superlattice reflections). It is argued that this exponent is due to a second order phase transition close to a tricritical point. Based on our detailed temperature dependent neutron diffraction studies, the phase diagram of sodium niobate is presented that resolves existing ambiguities in the literature.

# 1. Introduction

The past few decades have seen a tremendous flurry of research interest in perovskite materials due to fundamental interest and technological applications [1-6]. In this class of materials, alkaline niobates (like potassium sodium niobate and lithium sodium niobate with ultra-large piezoresponse comparable to $Pb(Ti_{1-x}Zr_x)O_3$ (PZT)) have evoked considerable interest as the next generation eco-friendly lead-free piezoceramics [3-6]. One of the end members, $NaNbO_3$ is a well documented antiferroelectric which finds applications in high density optical storage, enhancing non-linear optical properties, as hologram recording materials, etc [1, 3-6]. An understanding of its crystallographic structures and phase transitions is of fundamental interest due to its functional properties such as enhanced ferroelectric and piezoelectric response.

$NaNbO_3$ exhibits an unusual complex sequence of temperature, pressure and particle size driven structural phase transitions [7-18]. Above 913 K, $NaNbO_3$ has paraelectric cubic phase with space group $Pm\bar{3}m$. On cooling, it undergoes a series of antiferrodistortive phase transitions ranging from paraelectric cubic phase to ferroelectric rhombohedral phase via intermediate paraelectric tetragonal, orthorhombic phases and antiferroelectric orthorhombic phase (see Fig. 1 of ref. 7). These temperature-induced structural phase transitions are driven by the so-called "soft-mode" mechanism. This concept is built on the assumption that the crystal gets unstable against particular Brillouin zone centre/boundaries phonon vibration modes [19]. The zone boundary phonon vibrations involve rotations of $NbO_6$ octahedra. The $M_3$ (q= ½, ½, 0) mode represents in-phase rotation of the adjacent layers (same direction $a^0a^0c^+$ in Glazer notation), while the $R_{25}$ (q= ½, ½, ½) mode represents the anti-phase rotation (opposite manner, $a^0a^0c^-$ in Glazer notation) [20]. In soft mode formalism, cubic to tetragonal ($T_2$ phase) and tetragonal to orthorhombic ($T_1$) phase transitions are driven by $M_3$ and $R_{25}$ phonon modes respectively. The diffuse x-ray scattering at R and M points into Brillouin zone is found to be strongly anisotropic and distributed along <100> reciprocal rods which can be reasonably explained by assuming contribution of the low-frequency zone boundary phonons along the line T (lying between M and R points). The phase transitions between the orthorhombic S, R and P phases [9-11] have not been investigated yet, and substantial enlargements of unit cell and complicated octahedral tilting are not understood.

Further, $NaNbO_3$ exists in ferroelectric phase at 15 K. The low temperature transition is driven by zone centre and zone boundary ($R_{25}$) modes. Our recent low temperature neutron diffraction studies provided unambiguous experimental evidence for the presence of coexisting ferroelectric phase ($R3c$) and antiferroelectric phase ($Pbcm$) over a wide range of temperatures (15 K- 275 K) in $NaNbO_3$ [8]. The ferroelectric ordering can be stabilized by applications of external field, chemical substitution and reduction of particles size even at room temperature [6,8]. The high-temperature phases of ferroelectric/antiferroelectric materials is of primary interest since the dynamics of the dipoles in these phases gives indication about the microscopic nature (e.g. order-disorder versus displacive) of the ferroelectricity itself.

In order to understand the complexity of the structural phase-transition sequence in NaNbO$_3$, first-principles calculations were carried out by Vanderbilt and coworkers **[14-18].** The calculations suggest presence of several competing structural instabilities with similar free energies. The authors included all the distortions involved in the observed structures of NaNbO$_3$ in their model calculations. The calculations predicted the coexistence of zone-center and zone-boundary phonon instabilities in cubic phase. However, they could not fully reproduce the observed sequence of structural phase transitions in NaNbO$_3$. The structural and dynamical properties, epitaxial strain, and anomalous LO-TO splitting of NaNbO$_3$ have also been reported.

In spite of the extensive experimental and theoretical studies as a function of temperature, there are numerous controversies surrounding the phase diagram of NaNbO$_3$, especially on the existence of ferroelectric ordering and correctness of the structures of the phases in different temperature regions. Darlington and Knight [9] have questioned the correctness of the structure at room temperature. They have shown that space group and cell dimensions for the high temperature R and S phases (stabilized between 633- 753 K and 753-793 K) as reported in the literature [9-11] are incorrect. Recently, on the basis of the work carried out using synchrotron X-ray diffraction, Yuzyuk *et al* [6 (a)] proposed a different sequence of structural phase transition in NaNbO$_3$. However, the correct crystal structures for the two high temperature R and S phases were still not determined.

The natures of antiferrodistortive phase transition in materials exhibiting perovskite structure have attracted enormous attention in the literature. Mainly, these phase transitions are driven by R point instabilities. We have carried out systematic neutron diffraction measurements, as a function of temperature from 300 K to 1075 K, to find the correct crystal structures for high temperature phases of NaNbO$_3$ and to understand their phase transitions behavior. To the best of our knowledge, studies have not been carried out to investigate the nature of antiferrodistortive phase transition driven by M point instabilities. We have also determined the effective exponent for the second order cubic to tetragonal phase transition using the intensity of the superlattice reflection and tilt angle. The critical exponent (β) is found to be nearly equal to a value β ≈ $^1/_3$. It is argued that this exponent is due to a nearly tricritical transition.

## 2. Experimental

The powder neutron diffraction experiments were carried out at the SINQ spallation source of Paul Scherrer Institute (Switzerland) using the high-resolution diffractometer for thermal neutrons (HRPT) [21] with wavelength 1.4940 Å. The data was collected under high intensity and high resolution mode with $\Delta d/d \geq 1.8 \times 10^{-3}$. All the data collections were carried out during heating cycles of the sample. The sample upto 1075 K was heated using a high temperature tantalum furnace. The structural refinements were performed using the Rietveld refinement program FULLPROF [22]. In all the refinements, the background was defined by a sixth order polynomial in 2θ. A Thompson-Cox-Hastings pseudo-Voigt with axial divergence asymmetry function was chosen to define the profile shape for the neutron diffraction peaks. Except for the occupancies of the atoms, all other parameters i.e., scale factor, zero correction, background and half-width parameters along with mixing parameters, lattice parameters, positional coordinates, and thermal parameters were refined. All the refinements have used the data over the full angular range of 4≤2θ≤ 164 degree; although in various figures only a limited range is shown for clarity.

## 3. Results and Discussion

### 3.1 Evolution of neutron diffraction as a function of temperature

Figure 1 depicts a portion of the powder neutron diffraction patterns of $NaNbO_3$ as a function of temperature in the range from 300–1075 K during heating cycle. We shall discuss the structures starting from the highest symmetry cubic phase that occurs at the highest temperature followed by the other phases occurring at lower temperatures. At the highest temperature (T= 1075 K), all the Bragg reflections present in powder diffraction patterns could be indexed as main cubic perovskite reflections. The reflection marked with arrow and labeled as F (2θ =35.1 °) in the diffraction pattern of cubic phase is from the furnace material (see Fig. 1(b)). Below 950 and 900 K, two new superlattice reflections appear at 2θ =35.0° and 37.0° (near to F) respectively. These superlattice reflections are associated with zone boundary M and R point instabilities, respectively. Further, below 810 K, an additional a set of superlattice reflections appear centered at 2θ =35.4° (marked with arrow and label as S5 in fig. 1(b)), and some of them diminish followed by enhancement of the intensity of super lattice reflection at 2θ =35.8° (marked with arrows and label as S3 in fig. 1(a)), which disappears below 680 K. Below 680 K, the super lattice reflections present at 2θ =35.0°, 35.8° and 41.7° (marked with arrows and labeled as S2, S3 and S4 in Fig. 1(a)) vanish and new super lattice reflections appear around 2θ =35.6° and 53.5° (marked with arrow and labeled as S1 in Fig. 1(a)) The intensity of these reflections increases gradually with decrease of temperature and reflections are present at room temperature.

NaNbO$_3$ has an antiferroelectric phase (*Pbcm*) at room temperature. In general, an antiferroelectric phase consists of two or more sub-lattice polarizations of anti-parallel nature, which in turn gives rise to superlattice reflections in the diffraction pattern. Superlattice reflections around 2θ =35.6$^o$ and 53.5$^o$ are one of the strongest antiferroelectric peaks and are marked with arrows and labeled as S1 in fig. 1(a). Thus, disappearance and reappearance of superlattice reflections in powder neutron diffraction provide unambiguous evidence for structural phase transitions in sodium niobate with temperature. Superlattice reflections, in different temperature regime, are also present at higher angles with prominent intensities. The details of symmetry and structure in different temperature regime are discussed in next section.

### 3.2 Rietveld Analysis of powder neutron diffraction data

#### 3.2 .1 Cubic (U) phase ( above T > 950 K)

The powder neutron diffraction patterns above 950 K are analyzed on the basis of an ideal cubic perovskite structure with the space group $Pm\bar{3}m$. The detailed structural parameters and goodness of fit for cubic phase of sodium niobate at 1075 K, as obtained from neutron diffraction data, are given in Table I. The fit between the observed and calculated profiles is satisfactory (figure 2(a)). Refinement with anistropic atomic displacement parameters gave better fit. Strongly anisotropic thermal vibrations of oxygen atoms, with the site symmetry 4/mmm, were observed. The anisotropy corresponds to the liberation of the NbO$_6$ octahedra. The general feature $U_{33}(O) \approx U(Nb)$ has been pointed out for cubic perovskite structures (ABO$_3$), which suggests structural phase transitions driven by octahedral librational M$_3$ and R$_{25}$ modes at low temperatures [23].

#### 3.2.2 Tetragonal (T$_1$) and orthorhombic (T$_2$) phases ((Temperature range 825 < T < 950 K)

The presence of superlattice reflection at 2θ =35.0$^o$, 37.0$^o$ below 950 and 865 K respectively in powder neutron diffraction patterns indicates the structural phase transitions from cubic phase. These superlattice reflections arise due to condensation of M and R zone boundary phonons and characteristics of tetragonal and orthorhombic phases respectively. Thus, for the temperature range (865 < T < 950 K) and (810 < T < 865 K) we refined the powder neutron diffraction data using tetragonal symmetry (space group: *P4/mbm*) and orthorhombic symmetry (space group: *Cmcm*) respectively. The result of Rietveld refinements are shown in Figure 2(b) and (c). We note that the contamination due to the furnace material decreased in lower temperature range.

#### 3.2.3   Orthorhombic S phases (Temperature range 755<T< 825 K)

The main feature of this phase is appearance of additional superlattice reflections beyond those in the T$_2$ phase at 2θ = 35.4$^o$ and 35.8$^o$. These superlattice reflections are marked with arrows and label as $S_5$ in fig. 1(b), could not be indexed using symmetry and space group of orthorhombic ($T_2$) phase. Absence of splitting in the main perovskite reflections and presence of new superlattice reflection suggest that multiplicity of cell is increasesed with respect to orthorhombic $T_2$ phase. The x-ray diffraction measurements by Ahtee *et al* [10] showed that S phase has orthorhombic symmetry with *Pnmm* space group and cell dimension 2×2×2 with respect to elementary perovskite cubic cell. However, Rietveld refinement of neutron diffraction data using the model proposed by Ahtee model [10] did not account for the superlattice reflection at $2\theta = 35.4$ degree.

In order to index additional superlattice reflection, we explored various possibilities and found that further multiplying of lattice parameter along [001] of the orthorhombic R phase (see next section) indexed all the reflections. Thus S phase has orthorhombic structure with space group *Pbnm* and cell dimensions $\sqrt{2}\times\sqrt{2}\times12$ with respect to elementary perovskite cell. This unit cell indicates that structure of S phase of NaNbO$_3$ is a modulated structure in which the lattice parameter along [001] is 2 times the $C_o$ lattice parameter of the R phase.

In the space group *Pbnm*, for the description of the crystal structure, we used the following axes: **$A_o$=-$a_p$+$b_p$ $B_o$=$a_p$+$b_p$**, and **$C_o$=12$c_p$**, were $a_p$, $b_p$ and $c_p$ correspond to the parent cubic phase perovskite. The orthorhombic unit cell contains 24 formula units. The detailed structural parameters and goodness of fit for S phase of sodium niobate at 770 K, as obtained from neutron diffraction data are given in Table II (A). The fit between the observed and calculated profiles is quite satisfactory and include the weak superlattice reflections as shown in Fig. 3(a). These superlattice reflections were also seen by Darlington and Knight [9(a)] using high resolution neutron powder diffraction data. To show the accountabilities of these super lattice reflections we have further plotted neutron data in term of d-spacing (Å) (see Fig. 3(b)).

### 3.2.4 Analysis of R phase (Temperature range 650<T< 770 K)

In the R phase, the characteristic superlattice reflections appear at $2\theta$ =35.0, 35.8, 37 and 41.7 degrees etc. We found that superlattice reflection at $\theta = 35.4^o$ disappeared followed by enhancement of the intensity of superlattice reflection at $2\theta$ =35.8 degrees of the S phase. The superlattice reflections in the R phase have (h, k/6, l); (h/2, k/2, l) and (h/2, k/2, l/2) indices (in terms of elementary pervoskite cell). The reflections appear due to condensation of DT (q= 0, 1/6, 0), M (q= ½, ½, 0) and R (q= ½, ½, ½) soft modes. The structure of R phase is not yet well investigated using neutron diffraction data. Sakowski has studied the structure of R phase; however the structural parameters were not refined [9(d)]. The author has shown that R phase has orthorhombic structure with space group Pnmm with cell dimension 2×6×2 times as large as that of the elementary perovskite cell.

The Miller indices of the superlattice reflections based on an elementary perovskite cell gives information about the nature of the octahedral tilts in the structure. Superlattice reflections with all-odd integered indices ("odd-odd-odd" i.e., "ooo" type in Gazer notation) and two-odd and one-even integered indices result from anti-phase (− tilt) and in-phase (+ tilt) tilting of the adjacent oxygen octahedral. This is due to structural phase transitions driven by softening and freezing of the phonons at R (q= ½ ½ ½ ) and M (q= ½ ½ 0) points of the cubic Brillouin zone, respectively. The presence of superlattice (h, k/6, l), (h/2, k/2, l) and (h/2, k/2, l/2) type reflections in powder neutron diffraction pattern confirm that these phase transition involve DT, M and R soft mode.

To find out the most probable space group we coupled DT, $M_3^+$ and $R_4^+$ irrep using the software package "ISOTROPY 2000 [24]" and obtained many space groups. The P phase of NaNbO$_3$ is orthorhombic (*Pbcm*) and undergoes first order phase transition to R Phase. Thus we imposed the constraints that R phase has higher symmetry than that of P phase *(Pbcm)*, primitive in nature and a subgroup of *Cmcm* and not follow group subgroup relation for *Pbcm*. This constraint reduced the number of possible space groups for the R phase. We find that the orthorhombic structure with space group *Pnma* (different setting of *Pbnm*) with cell dimensions √2×6×√2 with respect to elementary perovskite cell accounted all the reflection in the R Phase (see Figure 4). In the space group *Pbnm*, for the description of the crystal structure, we used the following axes: **A$_o$**=-**a$_p$**+**b$_p$** **B$_o$**=**a$_p$**+**b$_p$**, and **C$_o$**=6**c$_p$** , where a$_p$, b$_p$ and c$_p$ are elementary perovskite cell parameters. The orthorhombic unit cell contains 12 formula units. The detailed structural parameters and goodness of fit for R phase of sodium niobate at 680 K, as obtained from neutron diffraction data are given in table II (B). The fit between the observed and calculated profiles is quite satisfactory and includes the weak superlattice reflections. For comparison, we have also carried out Rietveld refinement using *Pnmm* space group with cell dimension 2×6×2. We found higher $\chi^2$ with the *Pnmm* space group compared to orthorhombic *Pbnm* space group although the former involved more number of parameters in the refinement.

### 3.2.5 Analysis of P phase (Temperature range 300 < T < 650 K)

The powder neutron diffraction data of P phase (300-650 K) include superlattice reflections which assume the indices (h, k, l/p), where h,k and l are integers, and p (four (4) in this case) gives the multiple along the [001] direction with respect to the equivalent cubic perovskite cell. These superlattice reflections are characteristic of the antiferroelectric phase. The powder neutron data have been analysed using orthorhombic structure with space group *Pbcm* (different setting of *Pbma*). The size of the orthorhombic unit cell is √2×√2×4 times as large as that of the high temperature cubic cell (elementary perovskite cell). Figure 5 (a) depicts the observed, calculated and difference profiles obtained by Rietveld analysis of the powder neutron data at 300 K. The fit between the observed and calculated profiles is quite good and include all the weak superlattice reflections. This confirms the orthorhombic structure with space group *Pbcm* of NaNbO$_3$ in temperature range from 300 to 650 K (Fig. 1). Based on synchrotron XRD diffraction studies, Yuzyuk *et al*. [6(a)] proposed that the P phase (from 250 to 650 K) consists of three phases: monoclinic (Pm) between 250 and 410 K, incommensurate between 410 and 460 K, and orthorhombic (Po) between 460 and 633 K. In view of this, we have collected data in small temperature interval and carried out careful Rietveld analysis. We have determined the structure using neutron diffraction, which offers advantage over x-ray diffraction, especially in the accurate determination of oxygen positions that are found to be responsible for subtle change in structures. However, we did not find any signature for the presence of above phase during Rietveld analysis of powder neutron data. The intensity of characteristic antiferroelectric reflection is found to decrease with increasing temperature and finally vanish at 680 K (see Figure 5 (b)).

### 3.3 Variation of lattice parameters

Variation of lattice parameters with temperature obtained from the Rietveld refinements is plotted in Fig. 6. The [110], [1 -1 0] and [001] directions of the cubic phase correspond to the [100], [010] and [001] directions respectively in all the orthorhombic P (space group: *Pbcm*), R and S (space group: *Pbnm*) phases, and tetragonal (space group: *P4/mbm*) phases. On the other hand, the lattice vectors of the orthorhombic T$_1$ Phase (space group: *Cmcm*) coincide with those of the original <100> cubic phase vectors. For the sake of easy comparison with the corresponding cell parameters of the various phases of sodium niobate, we have plotted equivalent elementary lattice parameters instead of lattice parameter corresponding to the different phases in Fig. 1. It is evident from the figure that in the antiferroelectric orthorhombic phase (P phase), the a$_p$ and c$_p$ parameters increase while the b$_p$ parameter decreases with increase in temperature. The discontinuous jump of lattice parameters at 633 K suggests a first order phase transition. Around 680 K, all the lattice parameters come close to each other and increase continuously with increase of temperature. Absence of abrupt change in the lattice parameters at the R to S phase transition temperature, suggests that the phase transition is of second order in nature. As we further increase the temperature, a$_p$, b$_p$ and c$_p$ lattice parameter of the orthorhombic (*Cmcm* phase) phase increases and a$_p$ parameter approaches b$_p$ around 865 K. At this temperature, the sample undergoes an orthorhombic to tetragonal phase transition. Further as temperature increases, the c$_p$ lattice parameter decreases, and finally approaches to a$_p$ parameter, and the sample transform into the cubic phase. The continuous change in the lattice parameters reveals that the successive phase transformations from orthorhombic R to high temperature cubic phase are not of a strong first order in nature.

# 4 Phase Transitions

Distinct changes or vanishing of the intensity of superlattice reflections observed in temperature dependence of neutron diffraction patterns of $NaNbO_3$ clearly reveal that it undergoes a series of phase transition as a function of temperature, and consistent with previous observations from the literature [9-11]. Based on powder synchrotron x-ray diffraction studies, Yuzyuk *et al.* [6(a)] claimed existence of three phases over temperature range 250 to 633 K, namely: monoclinic (*Pm*) between 250 and 410 K, incommensurate between 410 and 460 K, and orthorhombic (Po) between 460 and 633 K. We are able to refine all the diffraction pattern over 300 to 633 K with orthorhombic phase and do not find any evidence of a phase transition over the temperature range. However, it may be noted from Figure 7 that around 500 K, the plots of lattice parameter $a_p$ and $c_p$ cross each other. The microscopic origins of various phases that stabilized over 10 to 1075 K are discussed below.

Recently, Izumi *et al* carried out detailed inelastic neutron scattering study in the cubic phase of $NaNbO_3$. The measurements show gradual softening of transverse acoustic (TA) phonon modes at the zone boundary M (½ ½ 0) and R (½ ½ ½) points, indicating the instabilities of the in-phase and out of phase rotation of the oxygen octahedral about the [001] direction. The softening of these modes suggests low-lying flat transverse acoustic phonon-dispersion relation along the zone-boundary M-R line (T-line). The sequence of phase evolution depends on the condensation sequence of the soft modes. The $R_{25}$ mode is three fold degenerate and the $M_3$ mode is non-degenerate. The triply degenerate $R_{25}$ soft mode could be treated as made up of three components corresponding to the rotations of the octahedra around three separate [001] axes. If only one component condenses at the transition point, the resulting structure would be tetragonal *I4/mcm*. On the other hand, coupled condensation of all the three components would give a rhombohedral *R3c* structure. Similarly, depending on the number of condensed $M_3$ phonons in the **q**-star (successively one, two, and three), the symmetry related second–order phase transitions from the cubic perovskite could be to $Pm\bar{3}m$ - *P4/mbm*- *I4/mmm*-*Im3* respectively.

In case of $NaNbO_3$, in contrast to other perovskites like $SrTiO_3$, $CaTiO_3$, $KNbO_3$ etc, $M_3$ phonon condenses first on cooling followed by the $R_{25}$ phonon on further cooling. The first structural transformation is from cubic $Pm\bar{3}m$ to tetragonal ($T_1$) *P4/mbm* structure doubling the unit cell in the plane perpendicular to the rotation axes though one $M_3$ mode condensation. Further, condensation of the $R_{25}$ phonon leads to the orthorhombic *Cmcm* ($T_2$) phase. This is the first example of the $M_3$ mode instability in the cubic to tetragonal phase transition in crystal of perovskite structure. Further, the observation of (1) low-lying flat transverse acoustic phonon-dispersion relation along the zone-boundary

M-R line and (2) strongly anisotropic diffuse scattering [ref] that is distributed along <100> reciprocal rods at R and M point, strongly indicate that the low- frequency zone boundary phonons lying between the M and R points also contribute to the transitions. These phonons play important role in stabilization of different (P, S and R) phases in $NaNbO_3$. The successive phase transition from orthorhombic *Cmcm* ($T_1$) to other phases may be explained as follows. Comparison of neutron diffraction patterns of orthorhombic phases (P, R, S and $T_2$ Phase) reveals that P, R and S phases have additional superlattice reflections with respect to orthorhombic $T_2$ (*Cmcm*) phase. We propose that these additional superlattice reflections arise due to condensation of the zone boundary phonons lying on the T (M-R) line (**q**= ½, ½, g) as a function of temperature. The orthorhombic structure of S, R and P phases result from condensation of phonon mode (**q**= ½, ½, g) with g= 1/12, 1/6 and ¼ respectively. In other words, the structures of these various orthorhombic phases result from modulation of the high symmetry cubic phase associated with phonon modes at **q**= (½, ½, g). Further, freezing of all the $R_{25}$ modes along with a zone-centre phonon stabilizes the low temperature ferroelectric rhombohedral phase.

The continuous variation of lattice parameters with temperature for the orthorhombic $T_1$, S and R phases suggests that the successive phase transitions are of second order in nature. However, the phase transition from orthorhombic R to P phase is of strong first order in nature. We recall that the space groups for orthorhombic $T_1$, and antiferroelectric P phases are *Cmcm,* and *Pbcm* respectively. Since, *Cmcm* to *Pbnm* follow group-subgroup relation while *Pbnm* to *Pbcm* does not, we propose that space group for both the orthorhombic S and R phases is *Pbnm* but with different unit cell modulation which accounts for all the superlattice reflections observed in neutron diffraction patterns.

The nature of antiferrodistortive phase transition in materials exhibiting perovskite structure has attracted enormous attention [25-27] in the literature. Mainly, these phase transitions are driven by R point instabilities. To the best of my knowledge, no attempt has been made to investigate nature of antiferrodistortive phase transition driven by M point instabilities. With this sprit, we have determined the effective exponent using the intensity of the superlattice reflection and tilt angle for sodium niobate of the tetragonal to cubic phase transition. The general relation between the order parameter (p) and the intensity (I) of these superlattice reflections is I $\alpha$ $p^2$. Figure 7 (a) depicts the temperature dependence of the integrated intensity of the (310) superlattice reflection (associated with M point instabilities) as a function of temperature. The solid line gives the least squares fit for the power law I$\propto$ $(Tc-T)^{2\beta}$ for which a critical exponent of $\beta\approx0.33$ $\pm0.03$ was determined. The variation in the tilt angles with temperature, calculated from the refinements in the tetragonal phase is illustrated in figure 7 (b). This behavior as seen in $SrTiO_3$, $CaTiO_3$, $Sr_xCa_{1-x}TiO_3$ and certain other perovskites [25-27] is typical of tricritical phase transitions. The critical exponent ($\beta$) for the cubic to tetragonal-phase transition, as obtained from the temperature variation of primary-order parameter, is found to be a value of 1/3. Thus, it is argued that this exponent is also a nearly tricritical transition.

Before we close, we would like to mention that by combining our present neutron diffraction data and earlier work [7], we have modified the phase diagram proposed by pervious workers [2], which was mainly based on x-ray diffraction studies. We now proceed to discuss the new features of our phase diagram given in Fig. 8. At high temperatures, the structure of $NaNbO_3$ is cubic. First principles calculations predicted the coexistence of zone-center and zone-boundary phonon instabilities in the cubic phase. On lowering the temperature, it undergoes successively transition to tetragonal ($T_2$ phase) and orthorhombic ($T_1$ phase) phases as a result of the condensation of zone boundary M and R point instabilities, respectively. Further, on lowering the temperature, appearance of additional superlattice reflection in orthorhombic S, R and P phases with respect to orthorhombic ($T_1$) phase are linked to freezing of zone boundary phonons along the line T (lying between the M and R points) with (q= ½, ½, g) with g = 1/12, 1/6 and ¼ respectively. The inelastic neutron scattering and x-ray diffuse scattering experiments also suggest the phonon instabilities along the T line. Finally, $NaNbO_3$ undergoes rhombohedral phase transition due to condensation of both zone centre and zone boundary ($R_{25}$) modes. It is also important to state that for the orthorhombic S and R phases, we have also explored the possibilities for the incommensurate/commensurate modulation using JANA software [28] as observed for solid solution of Ba or other perovskites [29-30]. We refined the diffraction data of the S and R phases with incommensurate modulated structures using JANA software; we find that the refined modulation vector does not vary significantly with temperature and the refined structure in both cases is quite close to the corresponding commensurate structure we reported above. Therefore, we do not consider the structures of the S and R-phase as incommensurate, which is also consistent with transmission electron study by Chen *et al* [13]. Single crystal diffraction data is highly desirable for the further confirmation.

## 5. Conclusions

In summary, we have established the structures and space groups of the technologically important sodium niobate up to 1075 K. Distinct changes or vanishing of the intensity of superlattice reflections, as observed in temperature dependence of neutron diffraction patterns, clearly revealed that $NaNbO_3$ undergoes a series of phase transition as a function of temperature, ranging from non-polar antiferrodistortive to ferroelectric and antiferroelectric in nature. These phase transitions are mainly driven by zone boundary phonon instabilities at R and M point and along the line T (connecting the M and R points). Additional superlattice reflections are found to appear at 680 K (R Phase) and 770 K (S

Phase), which could be indexed by using an intermediate long-period modulated orthorhombic structure whose lattice parameter are 3 and 6 times the lattice parameter of the $CaTiO_3$ type *Pbnm* structure along <001> directions. The correlation of superlattice reflections with the instability of zone boundary phonons is discussed. The discontinuous jump in lattice parameters at 633 K suggests that the transition from the orthorhombic P to R phase is of first order in nature, while other successive phase transitions are of second order. The phase diagram of $NaNbO_3$ has been reviewed based on our detailed temperature dependent neutron diffraction studies that resolved certain ambiguities in the literature.

**Table I:** Structural parameters of NaNbO$_3$ obtained by neutron diffraction at 1075 K. The cubic lattice parameter is 3.9507(2) Å in the space group $Pm\bar{3}m$. The unwritten thermal parameters U$_{ij}$ are 0(zero).

| Temperature = 1075 K (U phase) | | | | | | |
|---|---|---|---|---|---|---|
| Atoms | Positional Coordinates | | | Thermal parameter (10$^{-2}$Å$^2$) | | |
|  | X | Y | Z | U11(Å)$^2$ | U22(Å)$^2$ | U33(Å)$^2$ |
| Na | 0 | 0 | 0 | 7.32(9) | 7.32(9) | 7.32(9) |
| Nb | 0.5 | 0.5 | 0.5 | 1.96(4) | 1.96(4) | 1.96(4) |
| O | 0.5 | 0.5 | 0 | 7.40(8) | 7.40(8) | 1.94(6) |
| R$_B$= 1.79; R$_p$=6.89; R$_{wp}$=8.45; R$_{exp}$=6.82; $\chi^2$= 1.54 | | | | | | |

**Table II:** Structural parameters of NaNbO$_3$ obtained by Rietveld refinement of neutron diffraction at 770 (S phase) and 680 K (R phase) using Pbnm space group.

| (A) Temperature = 770 K (S phase) | | | | (B) Temperature = 680 K (R phase) | | | |
|---|---|---|---|---|---|---|---|
| Atoms Positional Coordinates and thermal parameter | | | | Atoms Positional Coordinates and thermal parameter | | | |
| X | Y | Z | B(Å)$^2$ | X | Y | Z | B(Å)$^2$ |
| Na1 0.5260(2) | -0.0073(7) | 0.0410(7) | 0.710(1) | Na1 0.5138(1) | 0.4984(1) | 0.2500 | 2.185(3) |
| Na2 0.5299(9) | -0.0315(7) | 0.1182(9) | 2.026(6) | Na2 0.9863(2) | 0.0109(3) | 0.0872(1) | 2.990(1) |
| Na3 0.5107(5) | -0.0146(5) | 0.2092(2) | 3.132(6) | Nb1 0.5000 | 0.0000 | 0.0000 | 1.013(1) |
| Nb1 0.0000 | 0.0000 | 0.0000 | 0.844(4) | Nb2 0.4981(1) | 0.0041(2) | 0.3323(2) | 0.868(1) |
| Nb2 0.0068(8) | 0.0065(2) | 0.0822(2) | 0.878(3) | O1 0.5315(1) | -0.0267(1) | 0.2500 | 1.512(2) |
| Nb3 -0.0100(5) | -0.0005(1) | 0.1680(6) | 1.022(4) | O2 0.0597(2) | 0.4824(2) | 0.0835(1) | 2.684(4) |
| Nb4 0.0017(3) | -0.0012(4) | 0.25000 | 1.082(3) | O3 0.7553(1) | 0.2442(3) | 0.0040(2) | 2.210(5) |
| O1 0.0520(4) | -0.0007(7) | 0.0415(1) | 3.225(3) | O4 0.7070(3) | 0.2932(1) | 0.3391(4) | 0.453(3) |
| O2 0.0368(4) | 0.0557(6) | 0.1243(1) | 2.636(4) | O5 0.7151(2) | 0.2885(3) | 0.6734(3) | 0.573(4) |
| O3 -0.0207(5) | 0.0005(4) | 0.2083(3) | 0.773(3) | | | | |
| O4 0.1990(7) | 0.2903(6) | 0.0028(6) | 0.001(1) | | | | |
| O5 0.2663(2) | 0.2358(1) | 0.0817(8) | 1.811(2) | | | | |
| O6 0.7290(1) | 0.7636(3) | 0.0827(1) | 1.977(2) | | | | |
| O7 0.2028(1) | 0.2851(1) | 0.1692(4) | 0.541(1) | | | | |
| O8 0.7992(2) | 0.6980(5) | 0.1692(3) | 1.062(2) | | | | |
| O9 0.2055(3) | 0.2737(5) | 0.2500 | 2.482(3) | | | | |
| O10 0.7792(5) | 0.7163(1) | 0.2500 | 0.668(3) | | | | |
| Lattice Parameters (Å) A= 5.5555(2) (Å); B= 5.5556(2) (Å) C= 47.1489(9) (Å); Volume = 1455.230(8) (Å)$^3$ R$_B$= 5.84; R$_p$=10.2; R$_{wp}$=10.4; R$_{exp}$=4.01 $\chi^2$= 6.77 | | | | Lattice Parameters (Å) A= 5.5459(1) (Å); B= 5.5505(1) (Å) C= 23.5229(3) (Å); Volume = 723.087(5) (Å)$^3$ R$_B$= 6.85; R$_p$=10.3; R$_{wp}$=10.35; R$_{exp}$=4.05 $\chi^2$= 6.62 | | | |

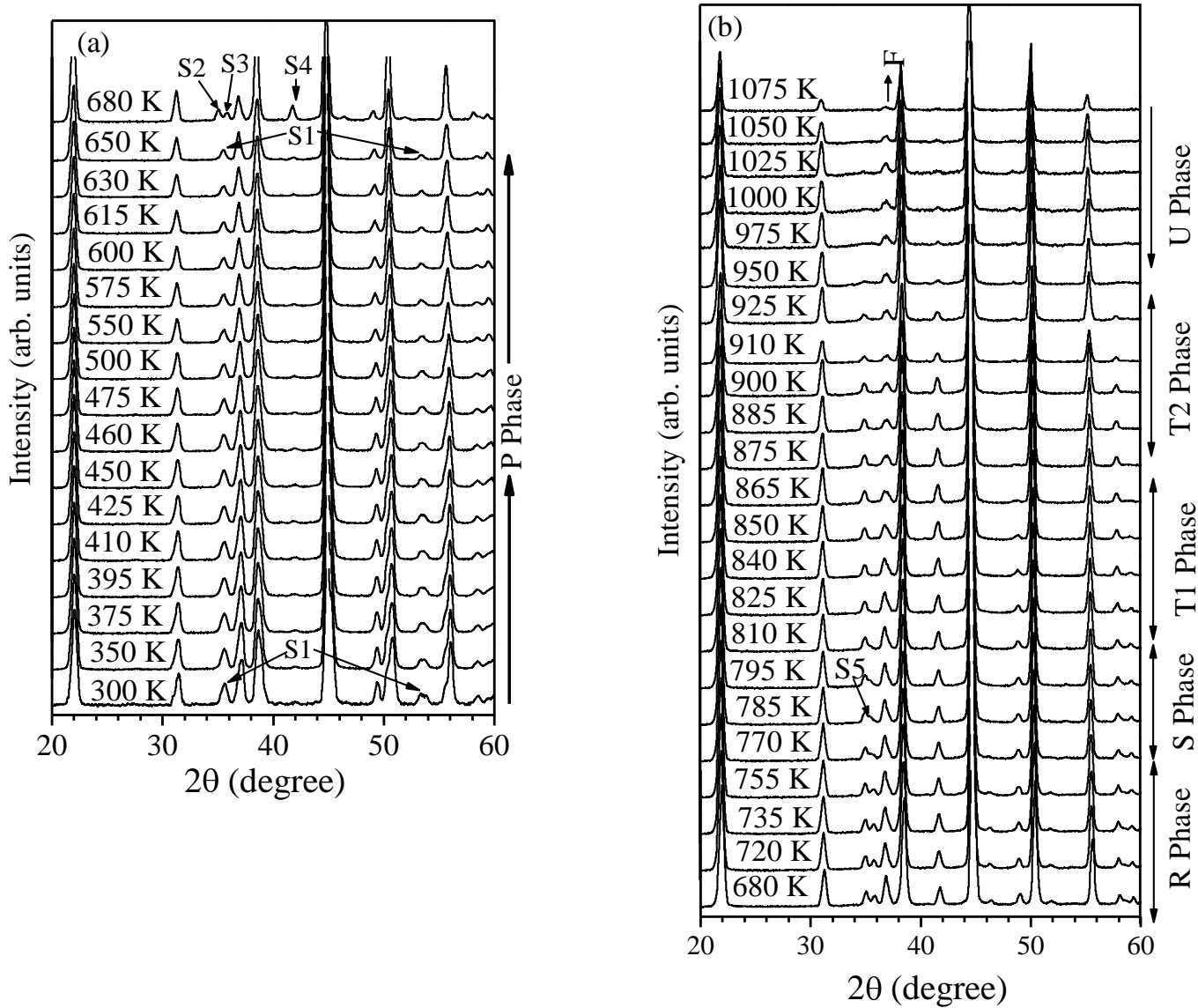

**Fig. 1** Evolution of the neutron diffraction patterns for NaNbO$_3$ as a function of temperatures. The characteristic superlattice reflections are marked with arrows and label as (S1, S2, S3, S4 and S5). Peak mark with F is due to the furnace materials.

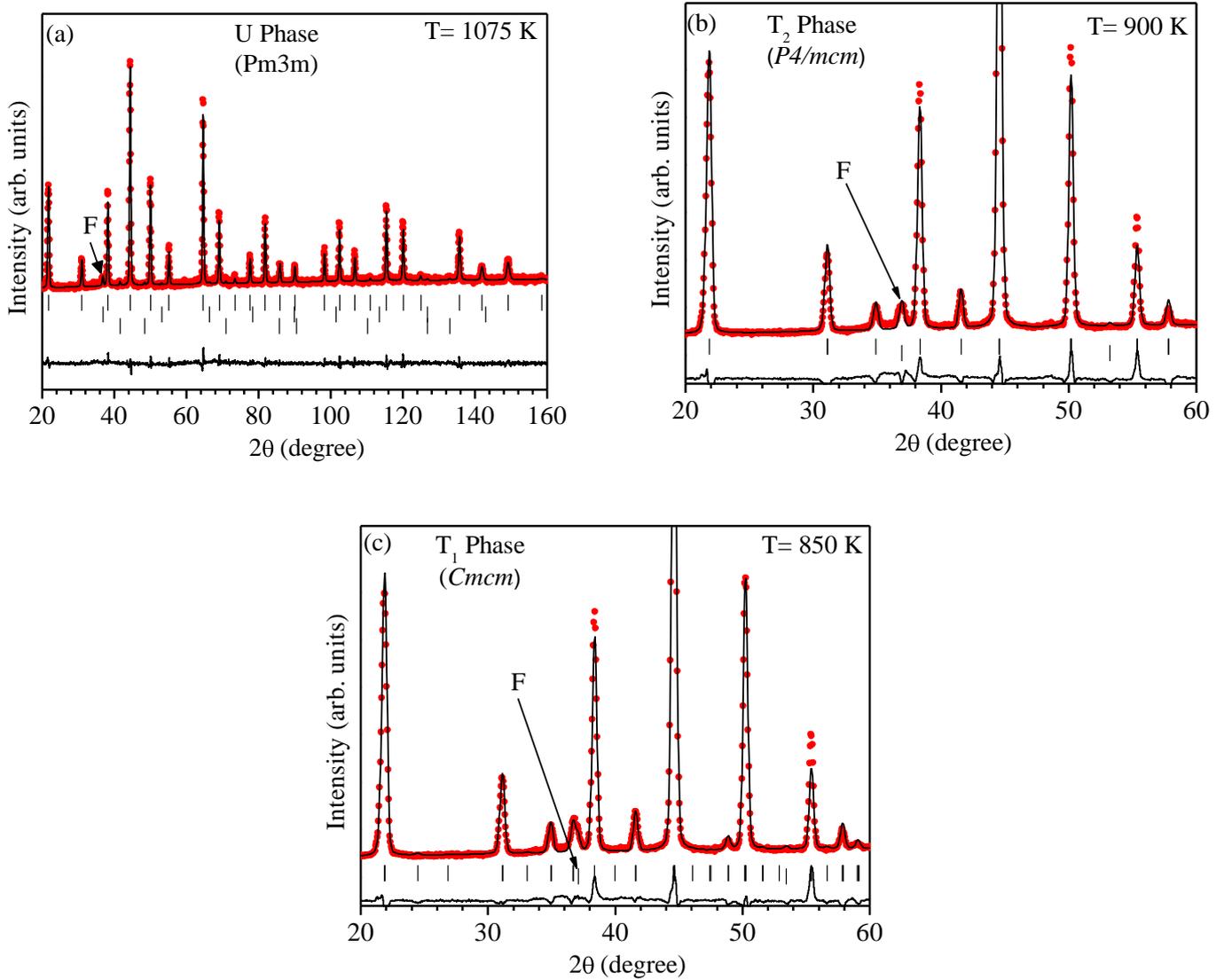

**Fig. 2** (color online) Observed (open circle), calculated (continuous line), and difference (bottom line) profiles obtained after the Rietveld refinement of $NaNbO_3$ using (a) cubic ($Pm\bar{3}m$), (b) tetragonal (*P4/mbm*) and (c) orthorhombic (*Cmcm*) phases. Peak mark with F is due to the furnace materials. In the cubic phase, middle and lower vertical tick marks above the difference profiles correspond to the furnace and thermocouple materials respectively.

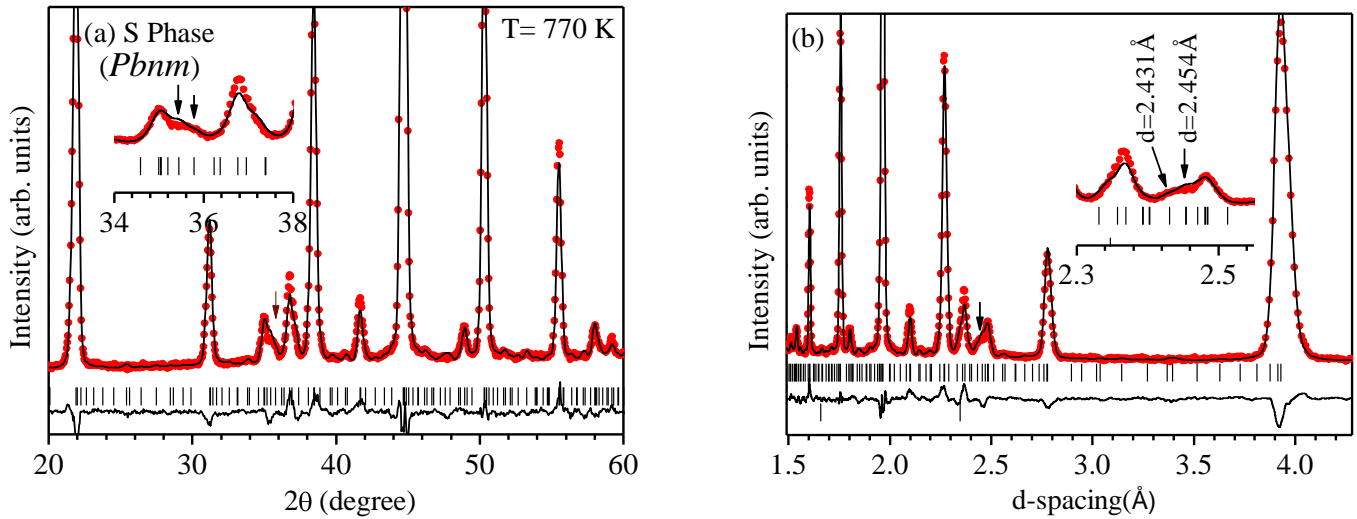

**Fig.3** (color online) Observed (open circle), calculated (continuous line), and difference (bottom line) profiles obtained after the Rietveld refinement of NaNbO$_3$ using long period modulated orthorhombic structure with space group *Pbnm* at 770 K. (b) In order to have an easy comparison with Darlington and Knight work [9(a)], the data has been plotted again in term of d-spacing (Å). Inset show the accountabilities of the characteristic superlattice reflections appeared in the S phase at 2θ = 35.4$^o$ 35.8$^o$ (in d-spacing term: d= 2.454 and 2.431 Å).

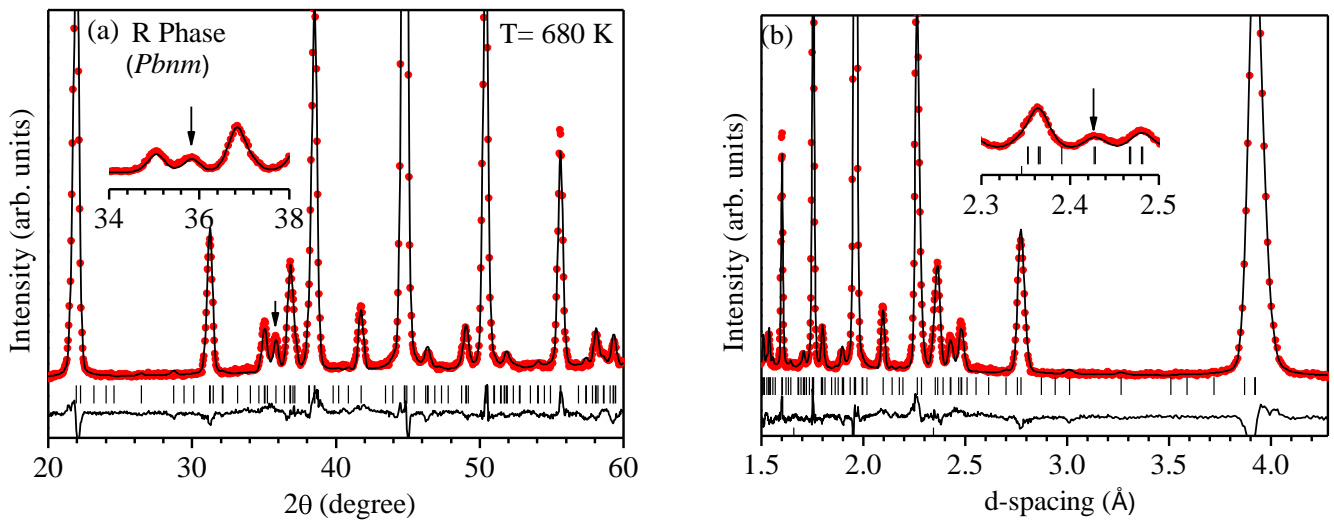

**Fig.4** (color online) Observed (open circle), calculated (continuous line), and difference (bottom line) profiles obtained after the Rietveld refinement of NaNbO$_3$ using orthorhombic Pbnm space groups at 680 K. The arrow shows the accountability of new superlattice reflection. (b) In order to have an easy comparison with Darlington and Knight work [9(a)], the data has been plotted again in term of d-spacing (Å). Inset show the accountabilities of the characteristic superlattice reflections appeared in the R phase.

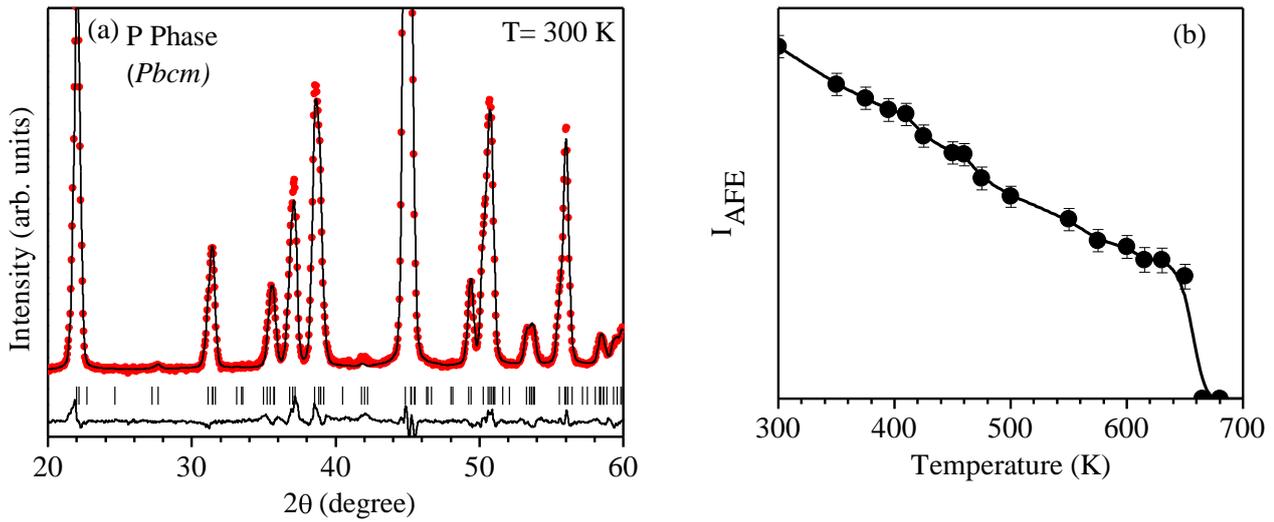

**Fig. 5** (color online) Observed (open circle), calculated (continuous line), and difference (bottom line) profiles obtained after the Rietveld refinement of $NaNbO_3$ using orthorhombic Pbcm space groups at 300 K. (b) Variation of the integrated intensity of one of the strongest antiferroelectric reflection appear around $2\theta = 35.6°$ with temperature.

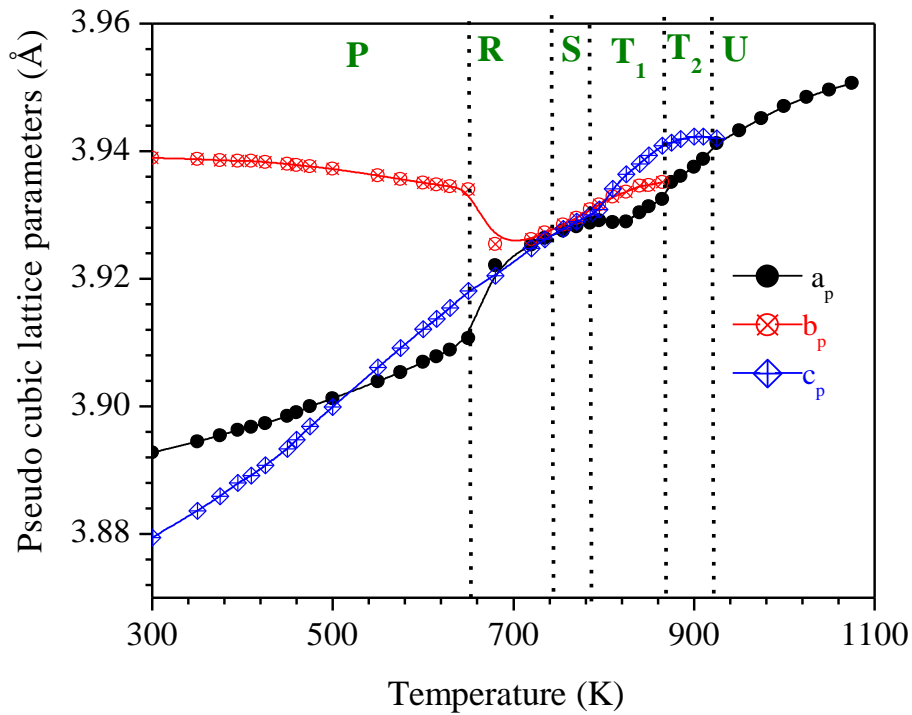

**Fig. 6** (color online) The variation of the pseudo-cubic lattice parameters as a function of temperature obtained after Rietveld refinement for $NaNbO_3$.

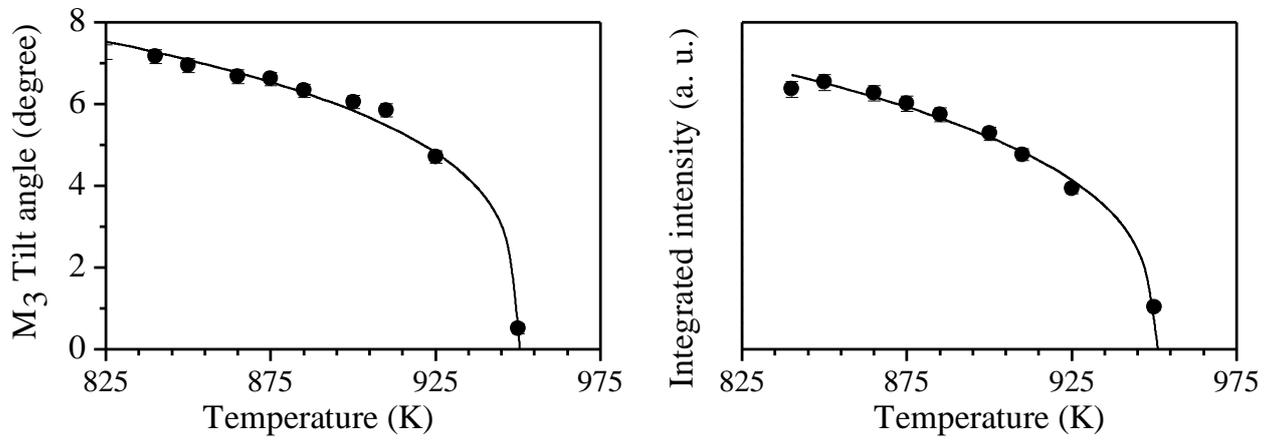

**Fig. 7** Variation of the tilt angle and integrated intensity of (310) reflection with temperature.

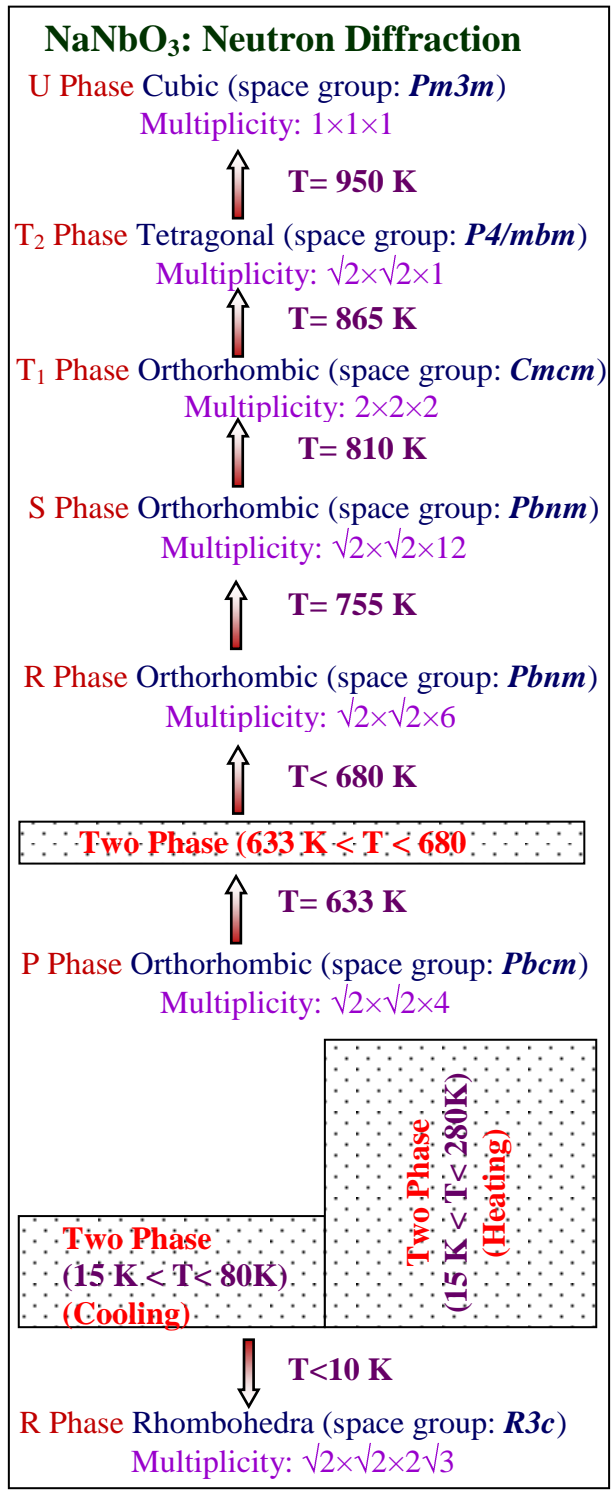

**Fig. 8** (color online) The phase diagram of NaNbO$_3$ as a function of temperature after neutron diffraction studies. The structure and space group of S and R phases correspond to our own present work; while the existence of two phase below room temperature are correspond to previous work.